\documentclass[aps,prl,showpacs,twocolumn]{revtex4-1}

\usepackage{graphicx}
\usepackage{dcolumn}
\usepackage{amssymb,amsmath}
\usepackage{natbib}
\usepackage{subfigure}
\usepackage{epstopdf}

\def\be{\begin{equation}}
\def\ee{\end{equation}}
\def\bea{\begin{eqnarray}}
\def\eea{\end{eqnarray}}

\def\begineq{\begin{equation}}
\def\endeq{\end{equation}}

\begin{document}

\title{Connecting flow structures and heat flux in turbulent Rayleigh-B\'enard convection}
\author{Erwin P. van der Poel$^1$, Richard J. A. M. Stevens$^{1}$,  and Detlef Lohse$^{1}$}
\affiliation{
$^{1}$ Department of Physics, Mesa+ Institute,  and J.\ M.\ Burgers Centre for
Fluid Dynamics,\\ University of Twente, 7500 AE Enschede, The Netherlands}
\date{\today}

\begin{abstract}
The aspect ratio ($\Gamma$) dependence of the heat transfer
(Nusselt number $Nu$ in dimensionless form)  in turbulent (two-dimensional) 
Rayleigh-B\'enard convection
is numerically studied in the regime $0.4 \le \Gamma \le 1.25$ for Rayleigh numbers
$10^7 \le Ra \le Ra^{9}$  and Prandtl numbers $Pr =0.7 $ (gas) and $4.3$ (water). 
$Nu (\Gamma )$ shows a very rich structure with sudden jumps and sharp transitions. 
We connect these structures to the way the flow organizes itself in the sample and explain
why the aspect ratio dependence of $Nu$ is more pronounced for small $Pr$. 
Even for fixed $\Gamma$ different turbulent states (with different
resulting $Nu$) can exist,    
between which the flow can or cannot switch. 
In the latter case the heat transfer thus depends on the initial conditions.
\end{abstract}

\maketitle

The physicists' view on fully developed turbulence has been dominated by Kolmogorov's seminal
work \cite{kol41a}, postulating universality of the small scales. Real-world flows however
have walls and boundaries. How do these geometric properties  
affect the global transport characteristics  of the flow such 
as heat or momentum transfer? The common view is that in the fully turbulent state, due 
to the large fluctuations of the system, the phase space is fully explored by the dynamics of the
flow and that universality implies an at most weak dependence on the boundary conditions, clearly
without any  jumps in global transport properties. 

Only recently the community became aware of the possibility of the coexistence of different
turbulent states, with first or second order phase transitions in between them. 
Examples include different turbulent states in von K\'arm\'an flow (``French washing machine''
\cite{cor10}), 
in magnetohydrodynamic turbulence in a closed system (von K\'arm\'an turbulent liquid 
sodium experiment) 
\cite{rav08},
in turbulent rotating  Rayleigh-B\'enard (RB) flow 
\cite{ste09}, or 
in turbulent rotating spherical Couette flow \cite{zim11}, and possibly even in 
high Rayleigh number turbulent RB flow in a cylindrical cell of aspect ratio $\Gamma = 1/2$
\cite{roc02,chi04,sun05a,xi08,wei10b,ahl11a,gro11} and $\Gamma=0.23$ \cite{roc10}.
That transitions between different flow states occur in closed flows
at low degree of turbulence has been well known and explored in the context of spatial-temporal
chaos and pattern formation, see e.g.\ \cite{bod00}, but that they also occur between
turbulent states clearly came as a surprise.

In this paper we will numerically 
show that different turbulent states with first and second order phase transitions 
in between them and even coexistence of different turbulent states
also exist in an even simpler
geometry, namely in a two-dimensional (2D) RB sample. 
The advantages of the 2D geometry are (i) that  a flow visualization is much easier,
(ii) that  the complicated axial and torroidal dynamics of 3D RB flow \cite{ahl09} 
does not complicate or even obscure the
flow field analysis, and (iii) that it is numerically cheaper so that a good resolution
in the aspect ratio $\Gamma$ (i.e. many different runs with different $\Gamma$) becomes
feasible. These features will enable us to reveal the connection between the flow
organization and the heat-transfer properties. 
However, we stress that these connections also exist in 3D RB flow, and working them out 
from  experimental measurements 
had been pioneered by Xia and coworkers \cite{xia}.

The main control parameter of this study will be the
aspect ratio $\Gamma$, whose effect on the turbulent RB flow has not yet been extensively studied,
with only a few theoretical and numerical studies \cite{gro03,chi06,bai10}. 
The main reason for this
short-coming is that it is experimentally very difficult to vary $\Gamma$
 in 3D cylindrical samples, as each $\Gamma$ requires a new sample. That is why 
in experiment the $\Gamma$ resolution has hitherto been insufficient to detect
transitions between different turbulent states. In their review Ahlers {\it et al.}
 \cite{ahl09} conclude, based on a small number of experiments with few 
different $\Gamma$ \cite{fun05,nik05,sun05e}, that the "weak $\Gamma$-dependence [of the heat transfer] suggests
an insensitivity to the nature of the large-scale convection''. We will show that in 
2D RB this is definitely {\it not} the case. 

One may argue that the flow phenomena in 
3D are different and richer, which is correct. However, various 2D RB flow simulations 
\cite{delu90,sch02,sug07,ahl08,sug09,joh09,sug10,zho10b}
have
revealed the value of this approach for a better understanding of turbulent RB convection
and the analogies to 3D RB flow. In ref.\ \cite{sug10} we could even reveal a one-to-one 
analogy between the flow organization in 
our 2D numerical simulations and in  experiments in a 
quasi-2D sample of the same aspect ratio $\Gamma = 1$. 

The code on which the results in this paper are based is a fourth order finite difference
discretization of the incompressible Oberbeck-Boussinesq equations.
It has been described and tested in detail in ref.\ \cite{sug09}. 
The velocity
boundary conditions on the walls are no-slip. The (relative) temperature is fixed at $\Delta/2$
at the bottom plate and $-\Delta/2$ at the top plate, with adiabatic side-wall conditions.
The grid resolution obeys the strict criteria formulated in ref.\ \cite{ste10}.
The imposed initial condition is a single roll state with a linear temperature profile.
The sample aspect ratio is $\Gamma = D/H$,
where $D$ is the sample width and $H$ the sample height.
Times are given in multiples of large eddy turnover times $t_E:=4\pi / \langle | \omega_c(t) | \rangle$, where $\omega_c$ is the center vorticity.
Next to $\Gamma$, the other control parameters are the Rayleigh number $Ra$ (the dimensionless
temperature difference between top and bottom) and the Prandtl number $Pr = \nu /\kappa$,
where $\nu$ and $\kappa$ are kinematic viscosity and thermal diffusivity, respectively.
The system responds with the heat transfer from bottom to top (in dimensionless form: Nusselt
number $Nu$) and the turbulence intensity (in dimensionless form: Reynolds number $Re$).

\begin{figure}[t!]
	\centering
\includegraphics[width=8.8cm,angle=-0]{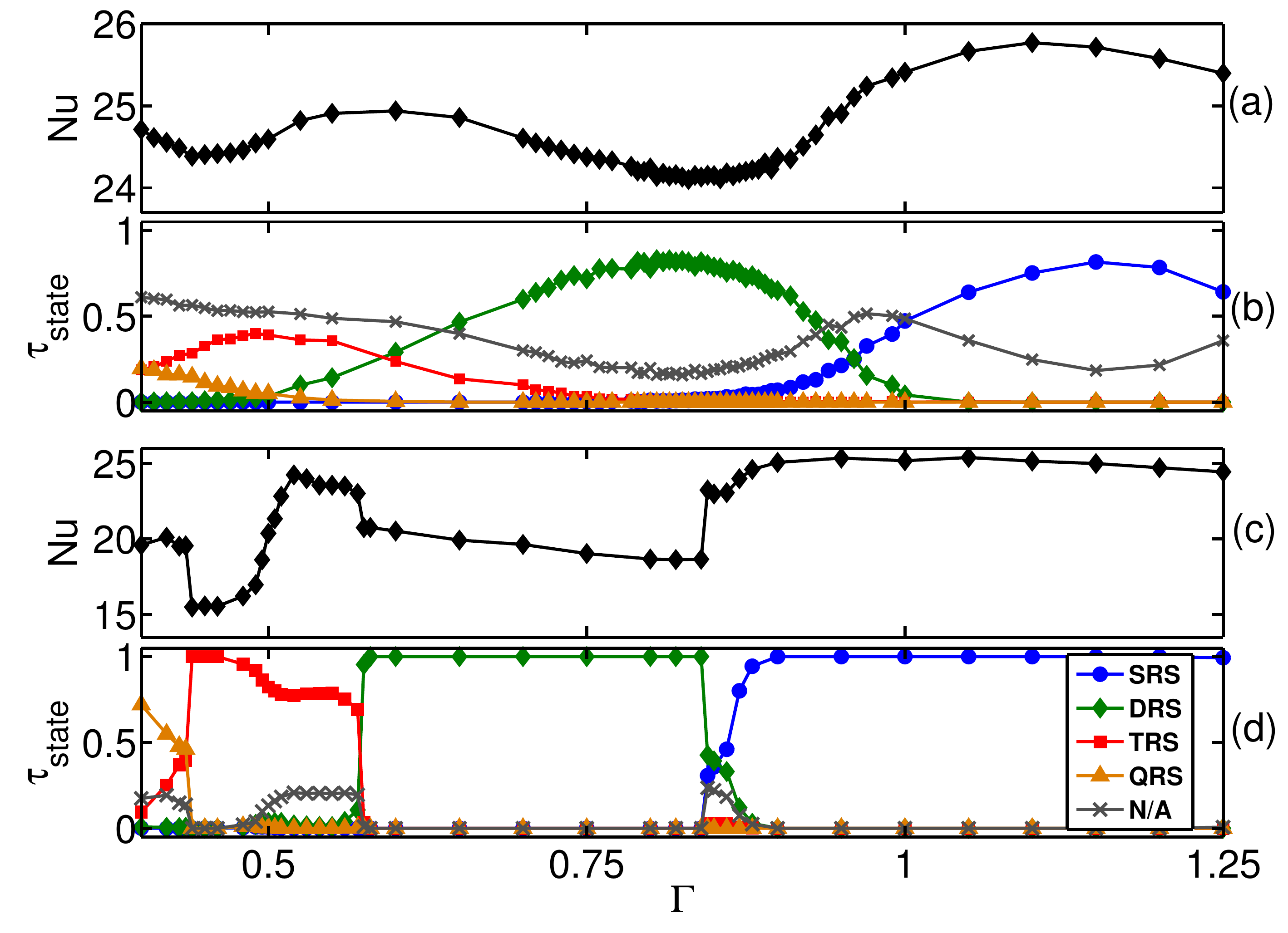}
    \caption{
(color online)
$Nu(\Gamma )$ for $Ra = 10^8$ 
and (a) $Pr= 4.3$  
and (c) $Pr=0.7$ (averaging time $>$ $10^3$ $t_E$).
The accompanying figures (b) and (d)  show the respective relative time the system is in 
the single roll state (SRS),
double-roll state (DRS),
triple-roll state (TRS),
and quadruple-roll state (QRS). Additionally, the extend of time that no roll state could be identified is indicated.
}
\label{fig1}
\end{figure}

\begin{figure}[t!]
    \centering
\includegraphics[width=8cm,angle=-0]{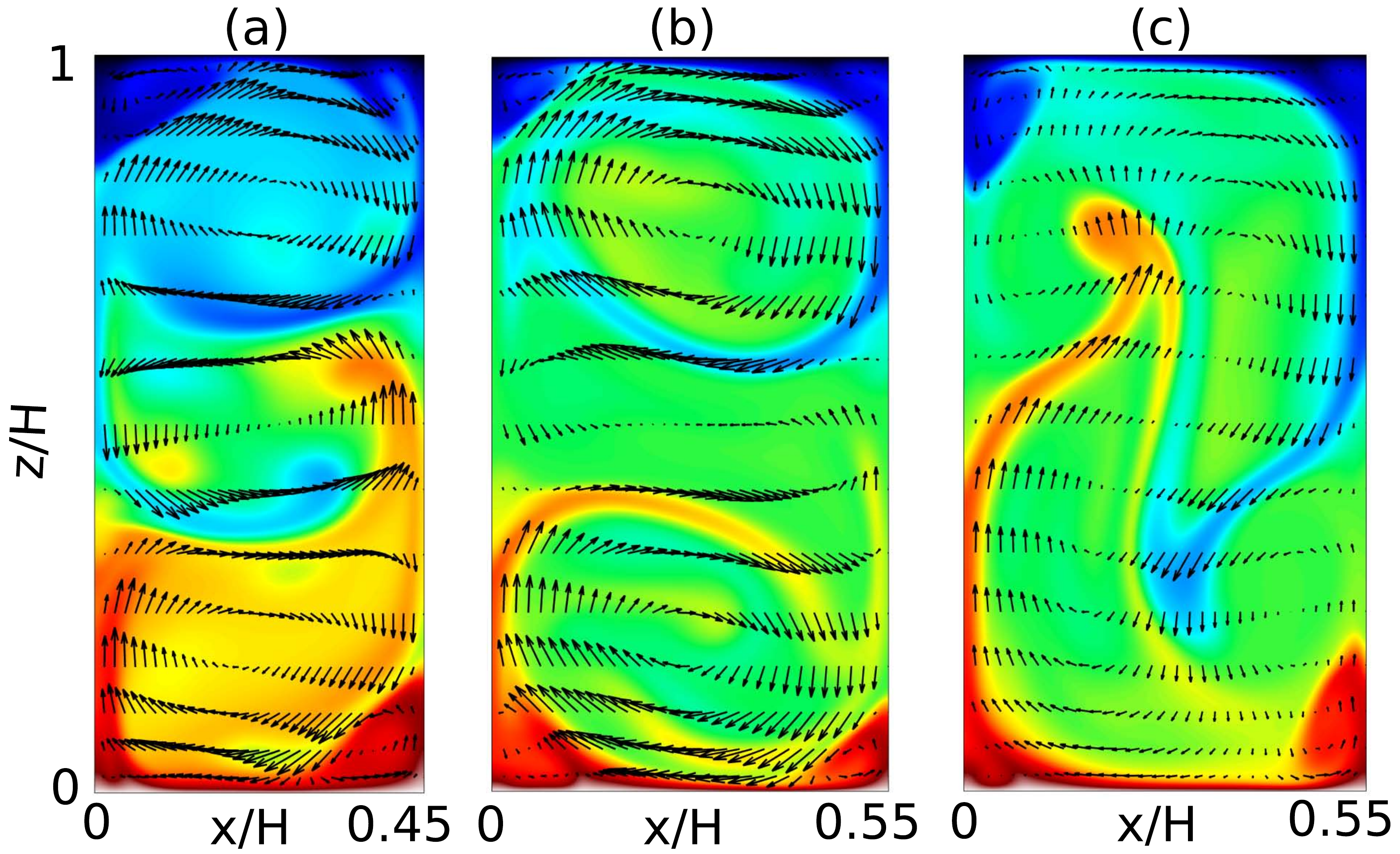}
\includegraphics[trim=0cm 0cm 0cm 0.5cm,width=4.25cm,angle=-0]{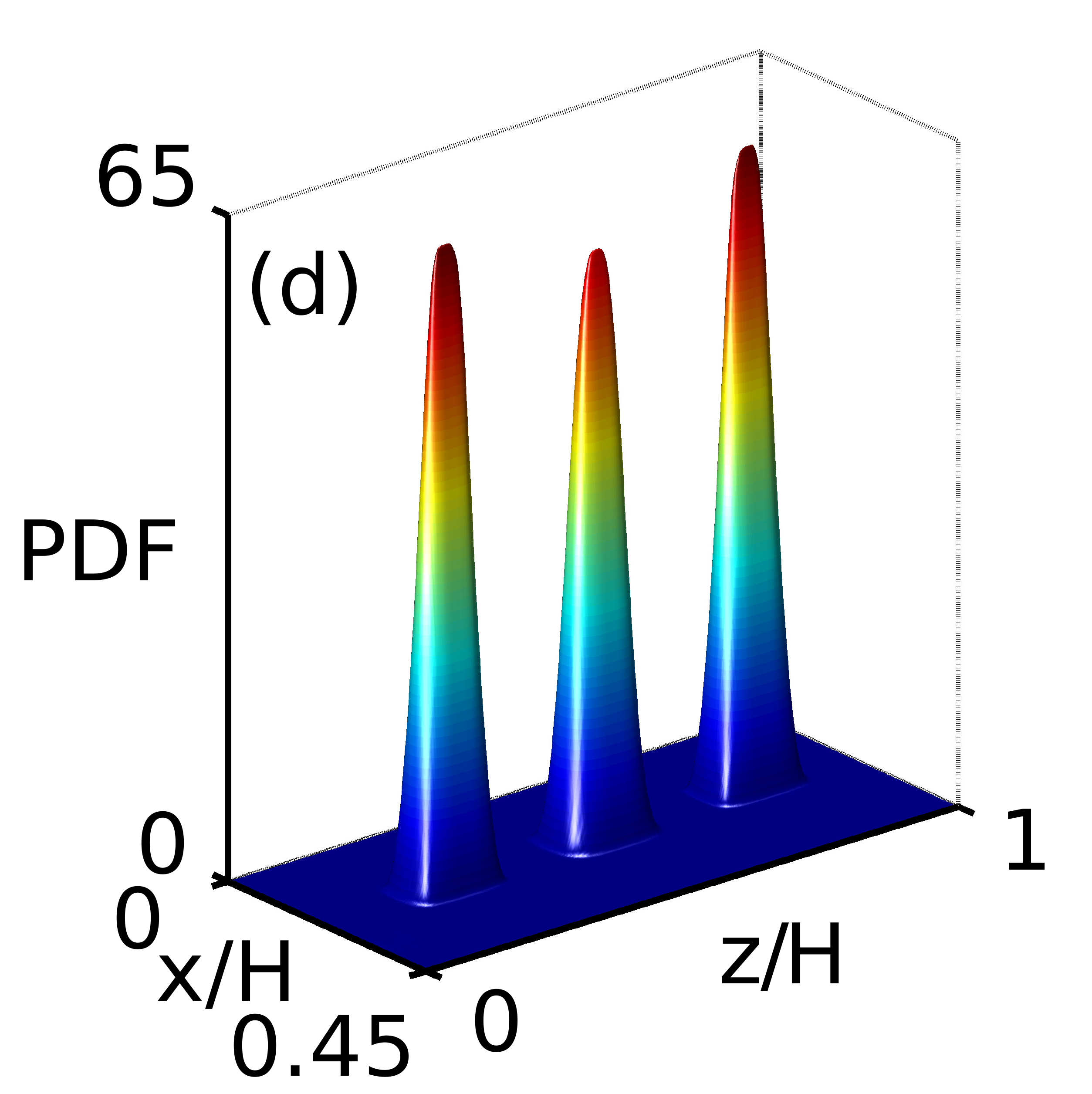}  
\includegraphics[trim=0.4cm 0cm 0cm 0cm,width=4.25cm,angle=-0]{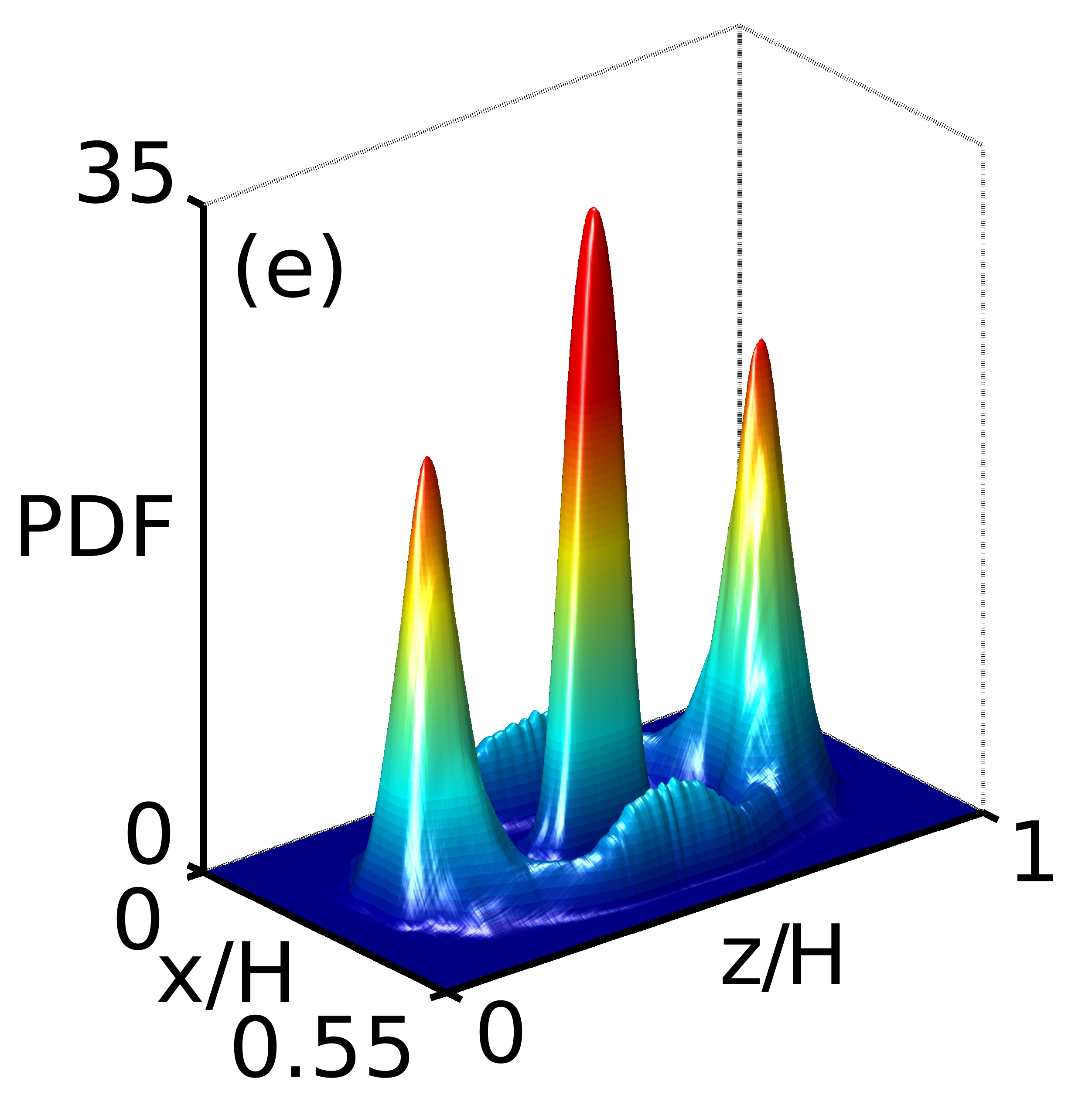}                       
    \caption{(color online)
Snapshots of the temperature field for $Ra = 10^8$ and $Pr=0.7$ for (a) $\Gamma = 0.45$ 
with the flow in a very stable TRS  
and (b,c) $\Gamma = 0.55$ with the flow in an unstable TRS. The top and bottom rolls in (b) grow until collision (c), which causes enhanced mixing and heat transport. Corresponding videos can be seen as supplementary
material. The positions of the roll centers are automatically tracked and their PDFs 
as function of (x,z) 
are shown in 
(d) for $\Gamma = 0.45$ and 
(e) for $\Gamma = 0.55$.  
}
\label{fig2}
\end{figure}

In fig.\ \ref{fig1} we show $Nu(\Gamma )$ for $Ra = 10^8$ and $Pr = 0.7$ and $4.3$,
resulting from a very large  averaging time of more than $10^3$ $t_E$ , for which we have checked statistical convergence.  
For the lower $Pr$ the curve shows variations in $Nu$ of up to 35\%, with abrupt jumps 
and sharp transitions. From visual inspection of flow movies we deduce that these jumps are associated with transitions in the overall flow organization
from one to two vertically stacked rolls (at $\Gamma \approx 0.9$) and from two to three rolls (at $\Gamma \approx 0.55$). We automize this flow analysis by 
measuring the zeros in the vorticity along the center vertical axis, i.e., the zeros
in  $\omega ( x= D/2, z)$. A state is not counted if its life time is shorter than $t_E$.
The results for the relative time in a certain state is shown in fig.\ \ref{fig1}. 
For $Pr = 0.7$ this analysis confirms that the transitions are rather sharp, with the 
coexistence of different roll states only in a small $\Gamma$-range. 
Next to the jumps, relatively sharp transitions in $Nu(\Gamma )$ can be observed.
As an example we discuss the 35\% increase in $Nu$ from $\Gamma = 0.45$ to $\Gamma = 0.55$,
namely from $Nu=16$ to $Nu=24$. 
This coincides with a developing instability of the triple roll state (TRS): The sample
becomes too wide for three rolls so that the three-roll structure from time to time breaks down.
This breakdown shuffles warm fluid upwards and cold fluid downwards, leading to a
larger temperature gradient across the thermal boundary layers (BLs) and thus to increased heat
flux. Typical snapshots of the temperature field in this regime are shown in fig.\ \ref{fig2};
the movies can be seen in the supplementary material.

The enhanced vortex mobility at $\Gamma = 0.55$ as compared to $\Gamma = 0.45$ can also
be deduced from the positions of the vortex centers, which we identify by a non-iterative ellipse-fit \cite{hal98} coupled with a vortex criterium based on the non-zero imaginary part of the velocity gradient tensor's eigenvalues.
The 2D probability functions (PDFs) of these positions are visualized in fig.\ \ref{fig2}d 
(for $\Gamma = 0.45$) and fig.\
\ref{fig2}e (for $\Gamma = 0.55$).
While in the former case we see three pronounced sharp peaks, reflecting the stability 
of that flow configuration, in the latter case the peaks are smeared out,
showing the vortex mobility which contributes to the enhanced heat flux.  

\begin{figure}[t!]
    \centering
\includegraphics[width=8.7cm,angle=-0]{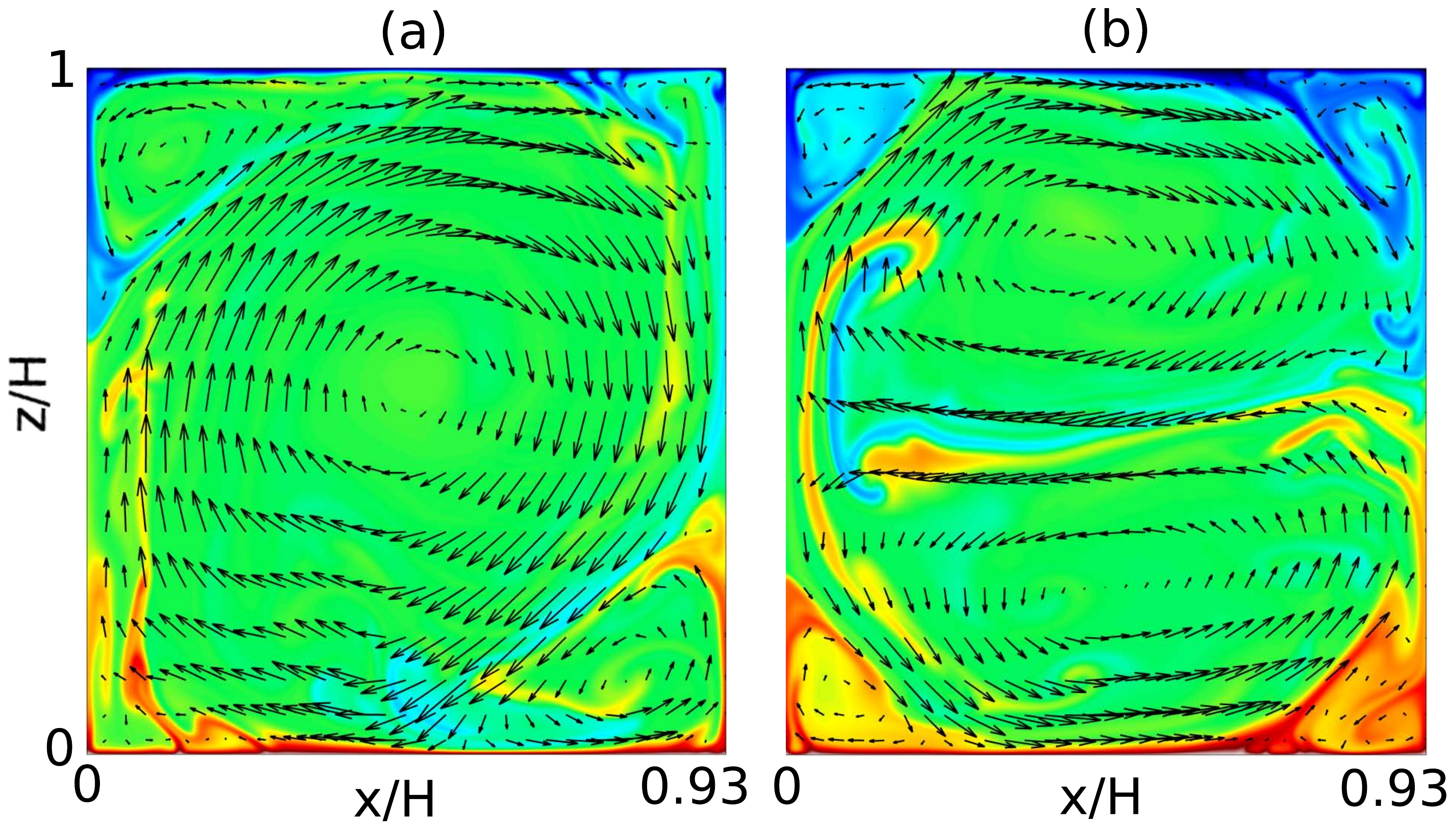}                            
    \caption{(color online)
Two temperature snapshots for $Ra = 10^9$, $Pr=4.3$,  and $\Gamma = 0.93$. 
The system is either in the SRS (a) or DRS (b). The corresponding movie is available online
as supplementary material. Heat transport efficiency differs between these states, see fig.\ \ref{fig4}.
}
\label{fig3}
\end{figure}

We now come to the case of 
 large $Pr=4.3$.
 Here 
the transitions in $Nu(\Gamma)$ 
are much smoother as compared to the $Pr = 0.7$ case
(compare figs.\ \ref{fig1}b and d) 
with wide $\Gamma$ regimes of coexistence between different
states, between which the dynamical evolution of the flow meanders, as e.g.\ 
seen from temperature field snapshots of fig.\ 
\ref{fig3}, the supplementary movie, and the time series of the Nusselt number.
For these larger $Pr$ the dynamics is much richer as the flow keeps on switching between
the different states. The reason for this lies in the plume dynamics: Whereas for 
$Pr=0.7$ the larger thermal diffusivity tends to smoothen out temperature differences 
between different flow parcels, for larger $Pr=4.3$ the plumes keep their thermal
energy for a longer time.  Thus the plumes, which feed the competing rolls as described in ref.\ \cite{sug10},  have a  longer life time,
 which leads to more lively roll dynamics. Therefore a 100\% dominance of a certain roll state, common for small $Pr$, does not occur. 

\begin{figure}[t!]
    \centering  
\includegraphics[width=8.0cm,angle=-0]{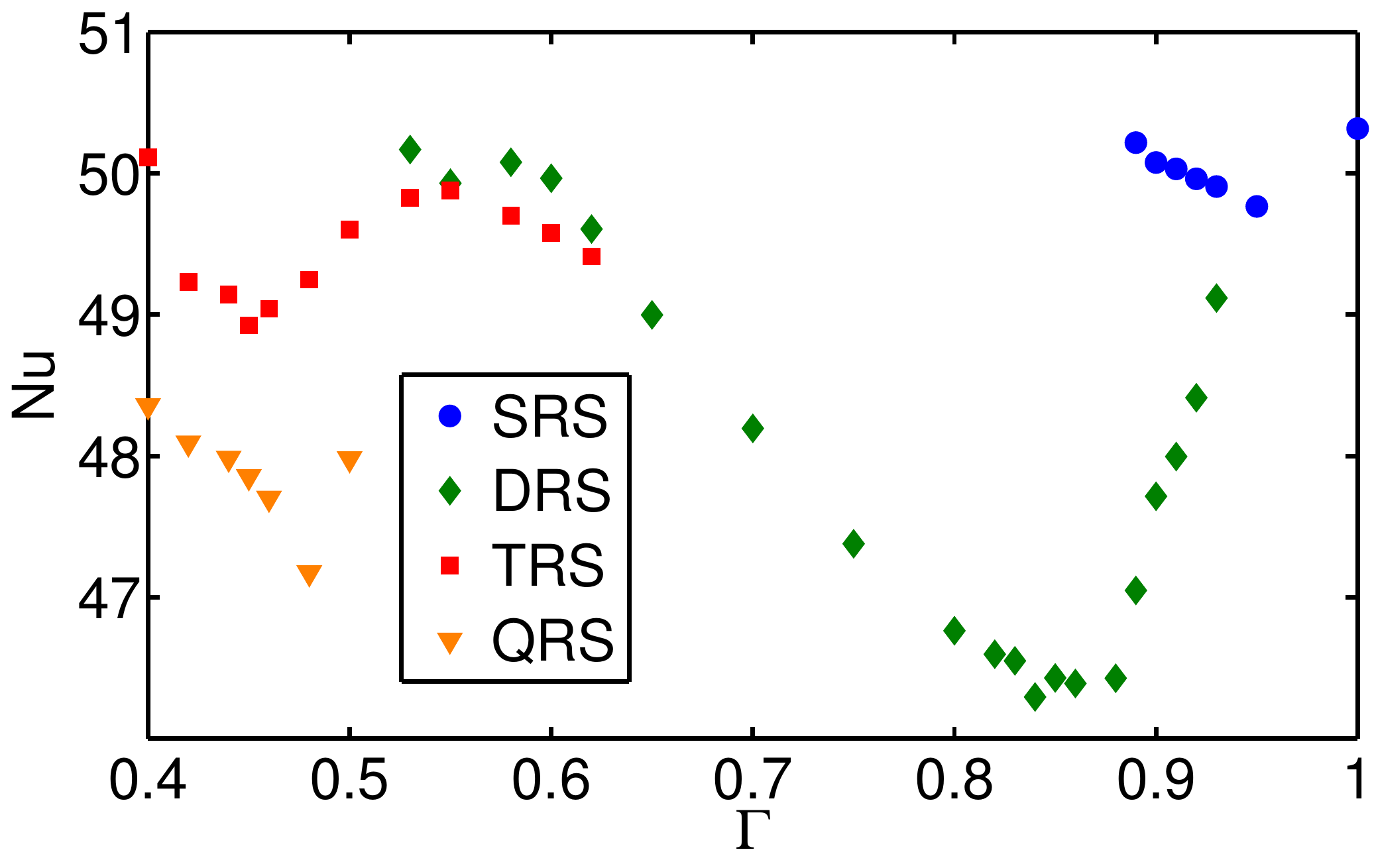} 
    \caption{(color online)
Conditionally averaged 
$Nu(\Gamma )$ for the various states SRS, DRS, and TRS, which can be identified at
$Ra = 10^9$ and  $Pr= 4.3$. 
For fixed $\Gamma$  
the flow jumps in between these different states.
}
\label{fig4}
\end{figure}

Based on the state analysis, the heat flux can be conditionally averaged on the SRS, DRS, or 
TRS. The results for these conditional heat fluxes are  shown in fig.\ \ref{fig4}. 
For the vertically arranging vortices it again holds that the stable states
with n rolls enable larger heat transfer than the stable states with n+1 rolls. For n=1 and n=2 
the difference is about 5\% for $Ra = 10^9$, $Pr=4.3$. Remarkably, that value
is similar to what one would expect for the 3D situation, based on an extrapolation towards $Ra = 10^9$ of the
$Nu$ measurements for the SRS and DRS in ref.\ \cite{wei11} (for $4\cdot 10^9 \le Ra \le 10^{11}$), which roughly gives 3\%. 

\begin{figure}[t!]
    \centering  
\includegraphics[trim=0.6cm 0.2cm 1.6cm 0cm,width=9.1cm,angle=-0]{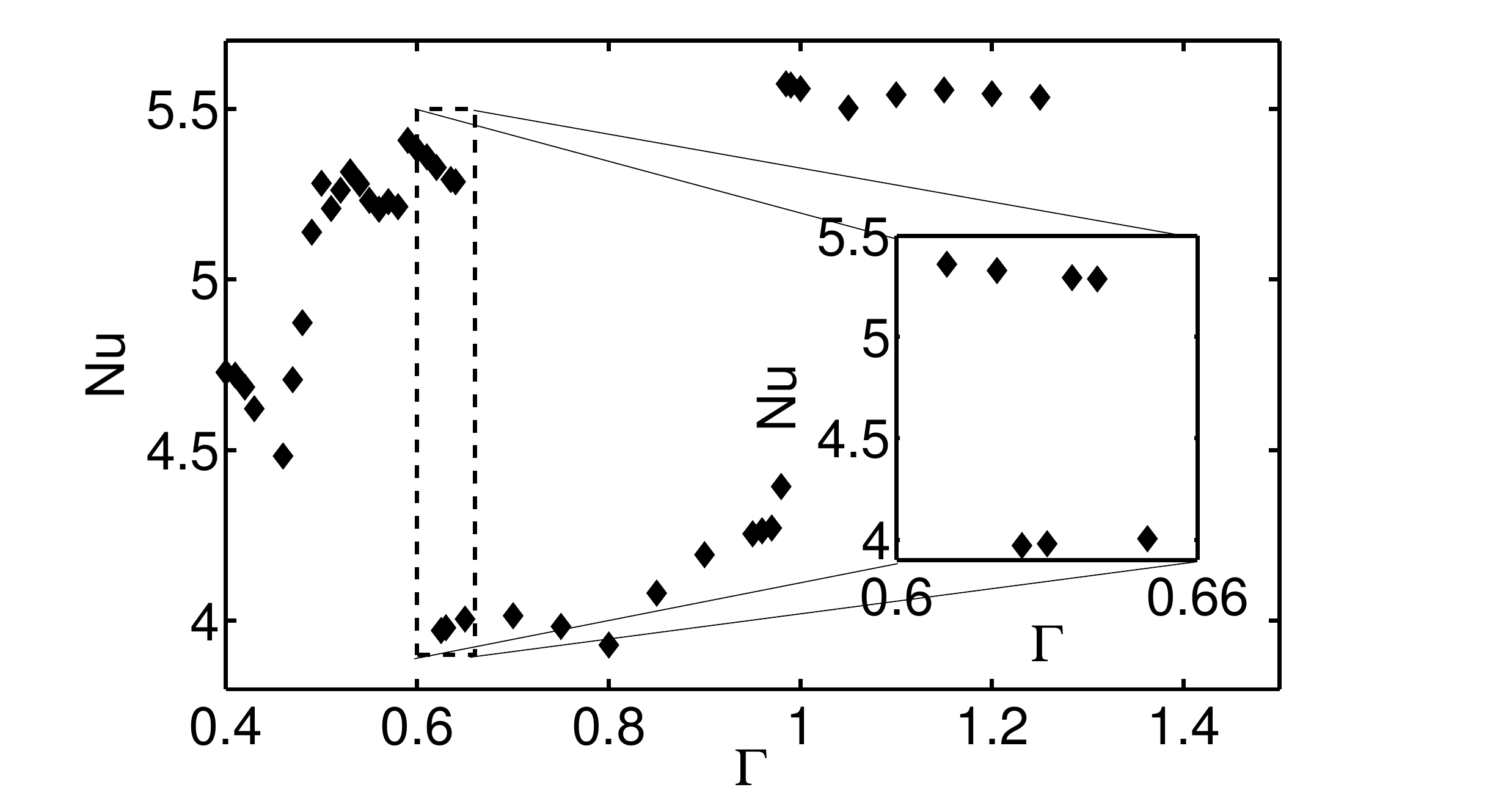}                       
    \caption{(color online)
$Nu(\Gamma )$ for $Ra = 10^7$ and $Pr=0.7$. For $\Gamma \approx 0.64$ the Nusselt number depends on the initial conditions, see text for details.
}
\label{fig5}
\end{figure}

We also found a case for which the final state and thus $Nu$ depend on the 
initial conditions of the flow field, see fig.\  \ref{fig5} for $Ra = 10^7$ and $Pr=0.7$, which according 
to  ref.\ \cite{sug07} is already turbulent.
The unstable TRS initial condition for $\Gamma= 0.63$ falls back to the DRS, whereas for  
initial DRS condition at $\Gamma= 0.64$ the flow remains in this DRS 
 (with lower  $Nu$) for longer than 3000 $t_E$. It remains to be seen whether this explicit 
dependence on initial conditions can also be found at larger $Ra$.

\begin{figure}[t!]
    \centering
 \includegraphics[width=8.0cm,angle=-0]{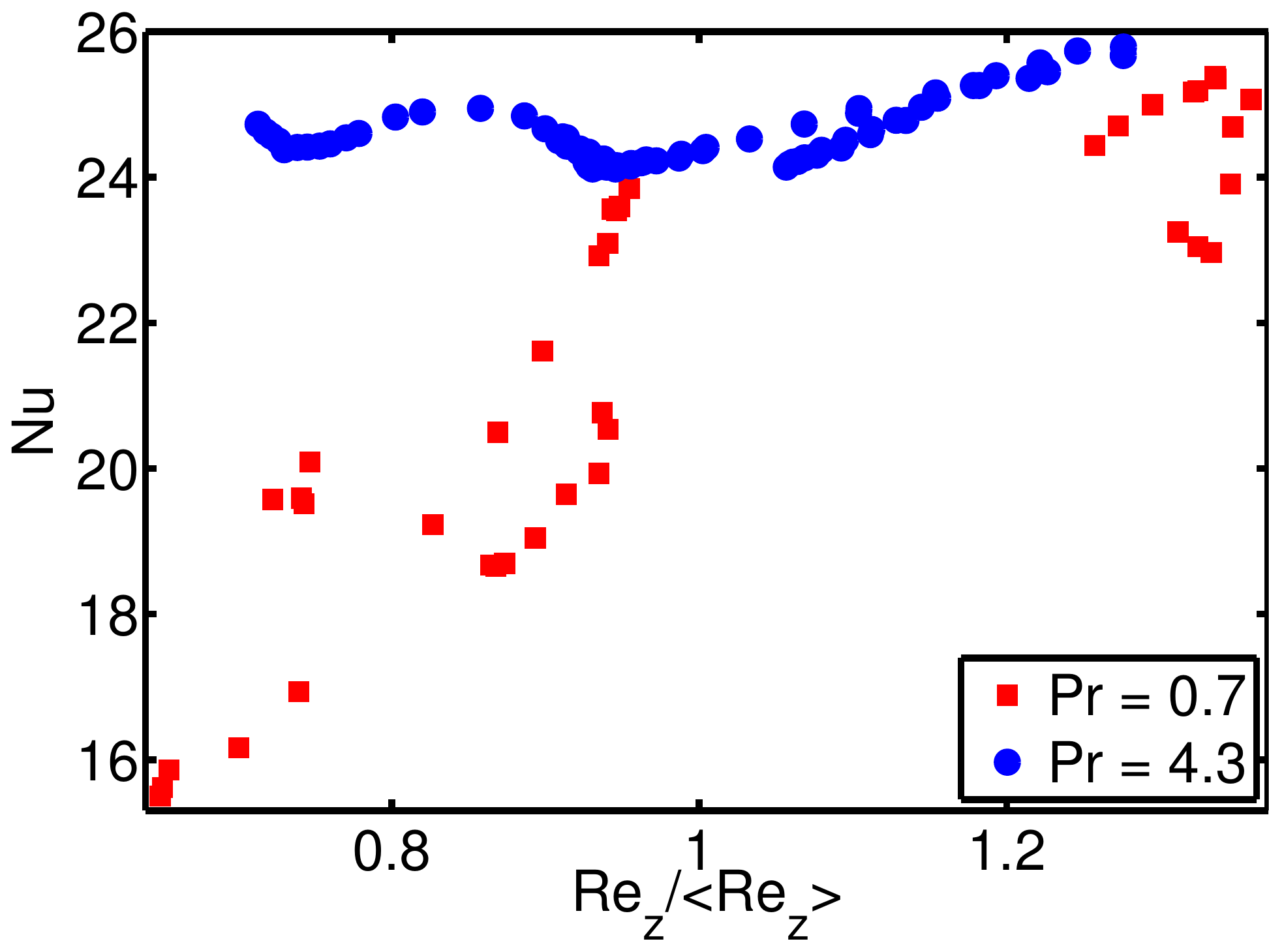}                       
    \caption{(color online)
$Nu (\Gamma) $ vs.\ $Re_z (\Gamma)/ \langle Re_z \rangle$ for $Pr=0.7$ (red squares) and 
$Pr= 4.3$ (blue bullets) at $Ra=10^8$.  
 }
\label{fig6}
\end{figure}

We finally address the question why $Nu$ is much more sensitive to the flow organization
at small $Pr$ as compared to large $Pr$, see again fig.\ \ref{fig1}. 
The first reason  for this finding is the organization of the thermal BL and kinetic BL: For large $Pr$ the thermal BL
is nested in the kinetic one. Thus modifications of the bulk flow are buffered by 
the kinetic BL and hardly lead to a different thermal BL thickness $\lambda_\theta$
 and therefore to hardly 
any change in $Nu \approx L / (2\lambda_\theta)$. In contrast, for low $Pr$ the thermal 
BL is thicker than the kinetic one and is therefore fully exposed to bulk flow modifications,
leading to a much stronger dependence of $Nu \approx L / (2\lambda_\theta)$ on the flow 
organization. The second reason for the larger sensitivity of $Nu$ at smaller $Pr$ is the larger thermal diffusivity, leading to a thermal smoothening of the plumes. 
As discussed above this results in  sharper transitions between different states. 
Our interpretation is supported by fig.\ \ref{fig6}, in which we plot
$Nu(\Gamma )$ vs  a normalized $Re_z(\Gamma)$, where $Re_z = u_{z,rms} L /\nu$, which represents
the bulk flow. For $Pr=4.3$ there is hardly any dependence of $Nu$ on
$Re_z / \langle Re_z \rangle$, whereas for $Pr=0.7$ this dependence is major. 
Here $\langle Re_{z}\rangle$ is the average Reynolds number of the shown $\Gamma$ range $0.4 \le \Gamma \le 1.25$.

In conclusion, 
we could clearly identify different
turbulent states, corresponding to different roll structures. 
It is remarkable that these features, which have been associated to lower Reynolds
and Rayleigh number flow in the regime of spatial-temporal chaos with much less dynamical
degrees of
freedom, survive for such 
high $Ra$. It seems that the turbulent states 'live' on low-dimensional structures.
The different turbulent states are associated with different overall heat transfer.
Transitions between the states -- either dynamically or as function of the 
control parameter $\Gamma$ -- thus imply 
jumps and sharp transitions in $Nu(\Gamma )$.
Based on the flow organization, we 
understand why these variations are larger for lower
$Pr$ and why the overall flow dynamics is richer at larger $Pr$, for which the plumes 
can keep their thermal identity for a longer time. 
The next step will be to push the present results to much higher $Ra$ in order to see
whether these features survive. Based on our analysis for $Ra = 10^8$ and $10^9$ 
we presume that this could be the case as the stability of the states and the variation in $Nu(\Gamma)$ is higher for $Ra = 10^9$ compared to $Ra = 10^8$ at $Pr = 4.3$. Additionally, large scale roll states have been distinguished up to $Ra = 10^{10}$ \cite{sug10} and can probably be found for even higher $Ra$. 

\vspace{0.5cm}
\noindent \textbf{Acknowledgements:} We thank K.\ Sugiyama for writing the first version
of the numerical code. The work was supported by
the Foundation for Fundamental Research on Matter (FOM) and the National Computing Facilities (NCF), both sponsored by NWO.



\end{document}